\documentclass[11pt,a4paper]{article}

\pdfoutput=1
\usepackage{jheppub}
\usepackage{graphicx,color,float}
\usepackage{hyperref}
\usepackage{multirow}
\usepackage{verbatim}
\usepackage{longtable}
\usepackage{amsmath}
\usepackage{amsfonts}
\usepackage{amssymb}
\usepackage{rotating}

\usepackage[normalem]{ulem}

\usepackage{siunitx}

\def\gsim{\raise0.3ex\hbox{$\;>$\kern-0.75em\raise-1.1ex\hbox{$\sim\;$}}}
\def\lsim{\raise0.3ex\hbox{$\;<$\kern-0.75em\raise-1.1ex\hbox{$\sim\;$}}}

\begin{document}


\title{Quark flavor violation and axion-like particles from top-quark decays at the LHC}

\author[a,b,c]{Kingman Cheung,}
\emailAdd{cheung@phys.nthu.edu.tw}
\affiliation[a]{Department of Physics, National Tsing Hua University,	Hsinchu 300, Taiwan}
\affiliation[b]{Center for Theory and Computation, National Tsing Hua University,	Hsinchu 300, Taiwan}
\affiliation[c]{Division of Quantum Phases and Devices, School of Physics, Konkuk University, Seoul 143-701, Republic of Korea}

\author[a,b]{Fei-Tung Chung,}
\emailAdd{feitung.chung@gapp.nthu.edu.tw}

\author[d,e]{Giovanna Cottin,}
\emailAdd{gfcottin@uc.cl}
\affiliation[d]{Instituto de F\'isica, Pontificia Universidad Cat\'olica de Chile, Avenida Vicu\~{n}a Mackenna 4860, Santiago, Chile}
\affiliation[e]{Millennium Institute for Subatomic Physics at the High Energy Frontier (SAPHIR), Fern\'andez Concha 700, Santiago, Chile}

\author[a,b]{Zeren Simon Wang}
\emailAdd{wzs@mx.nthu.edu.tw}

\date{\today}

\vskip1mm
\abstract{We study axion-like particles (ALPs) with quark-flavor-violating couplings at the LHC.
Specifically, we focus on the theoretical scenario with ALP-top-up and ALP-top-charm interactions, in addition to the more common quark-flavor-diagonal couplings.
The ALPs can thus originate from decays of top quarks which are pair produced in large numbers at the LHC, and then decay to jets.
If these couplings to the quarks are tiny and the ALPs have $\mathcal{O}(10)$ GeV masses, they are long-lived, leading to signatures of displaced vertex plus multiple jets, which have the advantage of suppression of background events at the LHC.
We recast a recent ATLAS search for the same signature and reinterpret the results in terms of bounds on the long-lived ALP in our theoretical scenario.
We find that the LHC with the full Run 2 dataset can place stringent limits, while at the future high-luminosity LHC with 3 ab$^{-1}$ integrated luminosity stronger sensitivities are expected.
}



\maketitle

%


\section{Introduction}\label{sec:intro}

In recent years, no signs of new, promptly decaying, heavy fundamental particles have been observed since the discovery of the Standard-Model (SM) Higgs boson at the LHC~\cite{ATLAS:2012yve,CMS:2012qbp}.
Thus, more attention has been given to non-traditional, possible forms of new physics (NP) beyond the Standard Model (BSM), such as light and feebly interacting particles~\cite{Agrawal:2021dbo,Antel:2023hkf}.
Such types of new particles are predicted widely in various NP models and are often long-lived, such that once produced they travel a macroscopic distance before decaying into either SM or other BSM particles.
In general, if such long-lived particles (LLPs) (see Refs.~\cite{Alimena:2019zri,Lee:2018pag,Curtin:2018mvb,Beacham:2019nyx} for recent reviews) are produced at the LHC, they can have escaped the past searches because these searches apply search strategies focusing on traditional signatures.
Given this possible reason of having missed discovery of new physics at the LHC, novel search strategies have been proposed and applied.
For instance, both ATLAS and CMS have published various types of LLP searches in recent years, targeting specific LLP collider signatures such as disappearing track~\cite{CMS:2023mny,ATLAS:2022rme}, displaced vertices and missing transverse momentum~\cite{CMS:2024trg,ATLAS:2017tny}, displaced vertex and a lepton~\cite{ATLAS:2020xyo}, displaced leptons~\cite{ATLAS:2020wjh,CMS:2021kdm}, and delayed or non-pointing photons~\cite{ATLAS:2022vhr}.
In addition, new dedicated experiments in the form of far detectors with a distance of $\sim 10-500$ meters away from the interaction points (IPs) have been under intensive discussion in the high-energy-physics community or even operated.
For instance, FASER~\cite{Feng:2017uoz,Ariga:2018uku,Feng:2022inv} is a relatively small and cheap cylindrical detector installed in the far forward direction of the ATLAS IP with a distance of 480 meters along the beam axis.
It has been running and collecting data during the LHC Run 3 phase, with early results already published~\cite{
FASER:2023zcr,FASER:2023tle}.
Other proposals include MATHUSLA~\cite{Chou:2016lxi,Curtin:2018mvb,MATHUSLA:2020uve}, MoEDAL-MAPP~\cite{Pinfold:2019zwp,Pinfold:2019nqj}, and CODEX-b~\cite{Gligorov:2017nwh,Aielli:2019ivi}.

Typical NP candidates of LLPs include sterile neutrinos, dark Higgs, electroweakinos in supersymmetry (SUSY) models, dark photons, and axion-like particles (ALPs).
Here, we focus on the ALPs and their phenomenology at the LHC (see for example Refs.~\cite{Bauer:2017ris,Alonso-Alvarez:2023wni,Ren:2021prq,Ghebretinsaea:2022djg,Baldenegro:2019whq,Florez:2021zoo} for some past studies in this direction).
The ALPs are pseudoscalar particles predicted in many UV-complete BSM models such as string compactifications~\cite{Cicoli:2013ana,Ringwald:2012cu}, supersymmetry models~\cite{Bellazzini:2017neg}, and Froggat-Nielsen models of flavor~\cite{Froggatt:1978nt,Alanne:2018fns}.
While they do not necessarily solve the strong CP problem in the SM as the QCD axions do with an extra $U(1)_{\text{PQ}}$ global symmetry breaking~\cite{Peccei:1977hh,Peccei:1977ur,Wilczek:1977pj}, their particular feature of having the ALP mass and the global symmetry breaking scale decoupled, allows for a rich phenomenology at various terrestrial experiments.
More concretely, we consider a scenario with a Lagrangian where the ALPs couple at tree level to quarks only, both diagonally and off-diagonally leading to flavor-changing-neutral-current (FCNC) interactions.
Such quark-flavor off-diagonal couplings can arise in various UV-complete models, either at tree level with DFSZ-type~\cite{Ema:2016ops,Calibbi:2016hwq,Arias-Aragon:2017eww,Bjorkeroth:2018ipq,delaVega:2021ugs,DiLuzio:2023ndz,DiLuzio:2017ogq} and KSVZ-type~\cite{Alonso-Alvarez:2023wig} axion models, or at loop level~\cite{Gavela:2019wzg,Bauer:2020jbp,Bauer:2021mvw,Chakraborty:2021wda,Bertholet:2021hjl}.
In particular, for these off-diagonal couplings, we restrict ourselves to those with the top quark and the up/charm quark.
The other off-diagonal couplings involving lighter quarks than the top quark are severely constrained by low-energy observables such as meson oscillations, and are therefore assumed to be negligible for simplicity in this work.
For the diagonal couplings, we include both up-type and down-type quarks in all generations.
We remain agnostic on a UV-completion origin for exactly this particular flavor structure and treat the ALP-quark couplings as independent parameters in this phenomenological work.
Existing phenomenological studies of quark-flavor-violating (QFV) ALPs can be found in e.g.~Refs.~\cite{Gorbunov:2000ht,Carmona:2021seb,Bauer:2021mvw,MartinCamalich:2020dfe,Carmona:2022jid,Li:2024thq}\footnote{See also Refs.~\cite{Blasi:2023hvb,Phan:2023dqw, Esser:2023fdo,Rygaard:2023dlx} for studies on the ALPs coupled to the top quarks only.}.

Owing to the large cross sections of the top-quark pair production at the LHC, we choose to study the rare decay of the top quarks into an ALP and an up/charm quark via the off-diagonal couplings.
The ALP is long-lived and decays hence with a displaced vertex (DV) dominantly into jets.
While the signal process studied here is almost identical to that in Ref.~\cite{Carmona:2022jid}, we approach the topic in a different manner.
Specifically, since the considered signal process involves multiple jets from both the prompt decays of the top quarks and the displaced decay of the ALP, we recast a recent ATLAS search for the same signature~\cite{ATLAS:2023oti} and thus reinterpret its exclusion bounds into the parameter space of the ALP model.

In Ref.~\cite{ATLAS:2023oti}, a ``search for long-lived, massive particles in events with displaced vertices and multiple jets'' was reported, at the center-of-mass (CM) energy $\sqrt{s}=13$ TeV with 139 fb$^{-1}$ integrated luminosity data collected during the LHC Run 2.
Exclusion bounds were obtained at 95\% C.L.~on long-lived electroweakinos originating in either strong or electroweak (EW) production channels.
Since the search required the LLP's mass to be larger than 10 GeV, we will hence focus on the ALP mass range around $[10, 100]$ GeV.

We now  discuss the current bounds on the ALPs coupled to quarks.
The existing limits mainly stem from low-energy precision measurements or collider searches; see e.g.~Refs.~\cite{Bauer:2021mvw,Carmona:2021seb}.
The primary low-energy precision observables include rare decays of kaons, $D$- and $B$-mesons, and $J/\Psi$, and are, for phase-space reasons, relevant only to the ALPs lighter than the bottom mesons.
For ALPs heavier than 10 GeV, collider searches have placed the primary constraints.
Firstly, Ref.~\cite{Esser:2023fdo} studies a theoretical scenario of the ALP coupled diagonally to a pair of top quarks only.
It considers existing bounds on the coupling from various sources including low-energy precision measurements and collider searches.
Specifically, for ALP mass between 10 and 100 GeV, the strongest upper bounds on $g_{33}/\Lambda$ (for the notation see Sec.~\ref{sec:model}) are derived by recasting an ATLAS search~\cite{ATLAS:2021hza} for ``events with two opposite-charge leptons, jets and missing transverse momentum'' and lie at $\sim 2\times 10^{-3}$ GeV$^{-1}$ (see Fig.~13 of Ref.~\cite{Esser:2023fdo}), in terms of an ALP signal process $pp\to t\bar{t} a$ with the ALP being detector-stable (with a corresponding signature of two top quarks +MET).
These constraints are relevant to our study since we assume homogeneous flavor-diagonal couplings of the ALPs to the quarks.
However, as we will see later in the numerical results, the ATLAS search performed in Ref.~\cite{ATLAS:2021hza} can, as shown in Ref.~\cite{Esser:2023fdo}, constrain the effective coupling, for $m_a\sim [10, 100$] GeV, orders of magnitude larger than those probed in this work by the ATLAS DV+jets search~\cite{ATLAS:2023oti}.

Secondly, Ref.~\cite{Carmona:2022jid} derives bounds on the ALP couplings to a top quark and an up/charm quark by recasting two searches:  i) a CMS search~\cite{CMS:2016uzc} for a single top quark plus jets (relevant to relatively more promptly decaying ALPs) and ii) an ATLAS search for top FCNC with a gluon mediator in the single top channel (which constrains detector-stable ALPs)~\cite{ATLAS:2015iqc}.
We follow Ref.~\cite{Carmona:2022jid} to recast these searches and reinterpret them in terms of our theoretical scenario, as detailed later in Appendix~\ref{sec:recast_single_top}.

Furthermore, a recent ATLAS search for displaced vertices consisting of jets~\cite{ATLAS:2024qoo} obtained exclusion bounds on the exotic top-quark decay branching ratio into an ALP and an up/charm quark as functions of the proper decay length of the ALP, for $m_a=40$ and 55 GeV.
We convert the top-quark decay branching ratios into the off-diagonal couplings responsible for the ALP production, and the ALP's proper decay lengths into the diagonal couplings mediating the ALP decays.
We find that this ATLAS DV search is weaker than the DV+jets search recast in this work by more than one order of magnitude on the ALP off-diagonal couplings across the whole sensitive range of the ALP proper decay length, if we confine ourselves to the strongest signal region of the DV+jets search.
Therefore, we choose not to present these bounds obtained in Ref.~\cite{ATLAS:2024qoo} in our numerical results.
At the end, we conclude that for the ALP mass range we study, current constraints on our model's parameters are all much weaker than the DV+jets search we study in all the relevant parameter regions, except the single top plus jet or MET searches to be explained later.

This paper is organized as follows.
In the next section we provide details of the model of the ALPs we consider, including the QFV terms.
Expressions of the signal decay widths of the top quark and the ALP are also given, together with the signal process studied in this work.
Then in Sec.~\ref{sec:experiment_simulation}, we describe the ATLAS search we recast, and elaborate on the simulation and computation procedures.
The numerical results of the LHC exclusion bounds and the projected future high-luminosity LHC (HL-LHC) sensitivity reach are presented and discussed in Sec.~\ref{sec:results}.
Finally, in Sec.~\ref{sec:conclusions}, we summarize the work and provide an outlook.
Additionally, in Appendix~\ref{sec:recast}, we detail the recasting procedure and results for the ATLAS DV+jets search, and in Appendix~\ref{sec:recast_single_top} we explain our method of reinterpreting single-top searches into our theoretical scenarios that can be relevant to promptly decaying and very long-lived ALPs.
\section{The ALP model with quark flavor violation}\label{sec:model}

We consider the following low-energy effective Lagrangian with ALP axial couplings to the quarks:
\begin{eqnarray}
  \mathcal{L}_{a,\text{ eff}} &=&  \frac{\partial^\mu a}{2\Lambda} \Big( \sum_{i=1,2,3} \, g_{ii}\  \bar{q}_i  \gamma_\mu  \gamma_5  q_i + \sum_{i,j=1,2,3}^{i\neq j} \, 
    g_{ij}\  \bar{u}_i  \gamma_\mu  \gamma_5  u_j   \Big )  \nonumber \\
    && + \frac{1}{2}(\partial_\mu a)(\partial^\mu a)  - \frac{1}{2}m_a^2 \, a^2, \label{eq:Lagrangian}
\end{eqnarray}
where $q_{i}$ labels either up-type or down-type quarks of generation indices $i$, and $u_{i,j}$ denotes the up-type quarks of generation $i,j$.
We assume that the $g$ couplings are real and positive, as well as symmetric in $i,j$ such that $g_{ij}=g_{ji}$ for $i\neq j$.
We do not include off-diagonal couplings for the down-type quarks in the theory.
$a$ denotes the ALP and $\Lambda$ is the effective cutoff scale (identified with the usual ALP decay constant $f_a$).
For numerical studies, we implement our model with FeynRules~\cite{Christensen:2008py,Alloul:2013bka} in the UFO~\cite{Degrande:2011ua} format.

We study the following signal process,
\begin{eqnarray}
    p p \xrightarrow{\text{SM}} t \bar{t}, \, (t \to W^+ b, W^+ \to jj), \,   (\bar{t} \to \bar{u}_i \,  a, a \xrightarrow{\text{disp.}} jj), \text{ with }i=1,2, \label{eq:signal_process}
\end{eqnarray}
and its charge-conjugated mode.
$jj$ denotes two jets including the $b$-quarks.
In particular, the ALP here is long-lived and decays to two jets with a macroscopic displacement from the IP.
For mediating the signal process, we assume all the diagonal couplings $g_{ii}$ with $i=1,2,3$, are non-vanishing and universal, and for the off-diagonal couplings, we consider only non-zero $g_{31}=g_{13}=g_{32}=g_{23}$ couplings for numerical studies.
In the rest of the paper, we sometimes refer to the $g_{3i}$ ($g_{ii}$) couplings as production (decay) couplings.

The top-quark decay width into the ALP and an up-type quark $u_i$ via $g_{3i}$ is,
\begin{eqnarray}
    \Gamma(t\to a u_i)&=&\frac{N_c}{384 \pi} \frac{g^2_{3i}}{\Lambda^2}\frac{m_a^2}{m_t}\Big(   \frac{(m_t^2 - m_{u_i}^2)^2}{m_a^2}    -  (m_t^2 + m^2_{u_i})  \Big)\nonumber \\
                            &&\times \sqrt{ \Big(  1 -  \frac{(m_a + m_{u_i})^2}{m_t^2}    \Big)    \Big(    1 - \frac{(m_a - m_{u_i})^2}{m_t^2}  \Big) }, \label{eq:Gamma_t2au}
\end{eqnarray}
and the same contributions from $g_{i3}$ are implied.
Here, $N_c=3$ is the number of QCD colors.
We then obtain the corresponding top-quark decay branching ratio (BR) by considering the experimentally measured value of the top-quark total decay width, $\Gamma_t=1.42$ GeV~\cite{ParticleDataGroup:2022pth}.

The ALP can decay to a pair of quarks with the same flavor $q_i\bar{q}_i$ via the diagonal couplings given in the Lagrangian, cf.~Eq.~\eqref{eq:Lagrangian}, and the corresponding decay width is given below~\cite{Bauer:2021mvw,Rygaard:2023dlx}
\begin{eqnarray}
    \Gamma(a\to q_i \bar{q}_i) = \frac{N_c m_a m_{q_i}^2}{8\pi}\frac{g^2_{ii}}{\Lambda^2}\sqrt{1-4 m_{q_i}^2/m_a^2},\label{eq:Gamma_a2qq}
\end{eqnarray}
applicable in the regime of perturbative QCD ($m_a\gtrsim 1$ GeV).
For ALP mass close to 100 GeV, the ALP decay BR into a pair of gluons via a triangle quark-loop with the same $g_{ii}$ couplings can be up to about 40\%~\cite{Carmona:2022jid}.
However, this channel leads to the same signature as those into light quarks, and its inclusion could only enhance the bounds on the production couplings to a rather minor extent (about 10\% in the large-decay-length limit).
Therefore, we choose not to include it in the computation for simplicity, and consequently, our numerical results are slightly conservative.

Similarly, the ALP can decay to a pair of photons via a triangle quark-loop.
We verified that for the $\mathcal{O}(10)$ GeV mass range of the ALP considered in this work, this loop-induced decay into two photons is suppressed by more than two orders of magnitude compared to that of the tree-level decay channel into a pair of quarks, and therefore we do not take into account this photon-pair channel in the numerical computation.
In addition, we ignore the ALP decays into a pair of quarks via a triangle loop which includes two quarks and a $W$-boson.

The non-zero off-diagonal couplings lead to 3-body or 4-body ALP decays that can be important if the production couplings are orders of magnitude stronger than the decay couplings:
\begin{eqnarray}
    a \xrightarrow{g_{3i}} \bar{u}_i \, t^* \to \bar{u}_i \, b \, W^{+(*)} \to \bar{u}_i \, b \, ( j \, j \text{ or } \, l^+ \, \nu),\text{ with }i=1,2,
\end{eqnarray}
and the charge-conjugate channel, where the $W$-boson can be either off-shell (4-body decays) or on-shell (3-body decays) depending on the ALP mass.
For ALP mass below roughly $m_W+m_b\sim 85$ GeV, the off-diagonal production couplings $g_{3i}$ can induce 4-body decays of the ALP which can be the dominant decay modes if the production coupling is larger than the decay coupling by at least around 4 orders of magnitude as we test with MadGraph5~\cite{Alwall:2011uj,Alwall:2014hca}.
Since the automatic calculation of the decay widths of these 4-body decay modes within MadGraph5 consumes too much computing resource, 
we do not include them in our computation and restrict ourselves to the parameter space where such contributions are unimportant.
For ALP masses above the $W$-boson threshold, the non-diagonal couplings can result in 3-body decays of the ALP and these modes are taken into account in our computation.
Moreover, flavor-diagonal ALP-quark couplings can also induce 4-body or 6-body decays of the ALP, via one or two off-shell quarks and hence $W$-bosons, respectively.
However, these decay channels are always largely suppressed and subdominant to the leading 2-body decays induced by the diagonal couplings, and are hence ignored.
\section{Experiment and simulation}\label{sec:experiment_simulation}

\subsection{The ATLAS DV+jets search}\label{subsec:search}

The ATLAS search~\cite{ATLAS:2023oti} focused on the final-state signature of DVs plus multiple jets, with the proton-proton collision CM energy $\sqrt{s}=13$ TeV and an integrated luminosity 139 fb$^{-1}$ collected during the Run 2 phase of the LHC.
The search employed two signal regions (SRs) called ``High-$p_T$ jet'' and ``Trackless jet''.
Both SRs started with a certain but different set of selections on the numbers of jets with various $p_T$ thresholds.
This step accounts for the event-level acceptance.
The two SRs then require that in the event there should be at least one vertex passing a list of vertex-level selections, including decaying inside the fiducial volume, having at least one displaced track with a transverse impact parameter larger than 2 mm, and DV invariant mass being larger than 10 GeV.
For the events that have passed both event-level and vertex-level acceptances, parameterized efficiencies, provided by the ATLAS collaboration\footnote{The relevant information including the materials for recasting can be found at the ATLAS HEPData webpage~\url{https://www.hepdata.net/record/ins2628398}.} to account for e.g.~multi-jet trigger, high-$p_T$/trackless jet filter, and material effects in the High-$p_T$-jet and Trackless-jet SRs separately, are applied in order to attain the final cutflow.

The ATLAS search considered a benchmark case of long-lived electroweakinos in the R-parity-violating supersymmetry (RPV-SUSY), in which these electroweakinos decay into quarks via baryon-number-violating $\lambda'' \bar{U}\bar{D}\bar{D}$ operators.
Two production channels of these electroweakinos are included.
The first is via strong interaction where gluinos are pair produced, which subsequently 
decay to the lightest neutralinos and jets.
The second is the direct electroweak production of a pair of electroweakinos including the lightest and the second lightest neutralinos as well as the lightest charginos.

We present the detail of the recasting procedures and validation results in Appendix~\ref{sec:recast}.
In this recast, we compare our cutflow efficiencies step by step in various benchmarks with different masses of the gluino or electroweakinos, as well as the latter's lifetimes, with those obtained with the ATLAS full simulation and with the ATLAS own recast.
For all the benchmarks, we achieve $\mathcal{O}(1)\%$ level agreement with the experimental, published results at the cutflow steps of the event and vertex acceptances, while, when the parameterized efficiencies are included, in some benchmarks deviations (``non-closures'') as large as $\mathcal{O}(10)\%$ are observed.

\subsection{Simulation and computation}\label{subsec:simulation}

With the Monte-Carlo (MC) simulation tool MadGraph5~\cite{Alwall:2011uj,Alwall:2014hca} and the UFO model that 
we have implemented, we generate one million signal events of the ALPs at the LHC (see Eq.~\eqref{eq:signal_process}) at multiple parameter points in a grid scan covering the production couplings, decay couplings, as well as the ALP mass.
The ALP decay width is automatically calculated within MadGraph5.

The generated signal-event files, in the LHEF~\cite{Alwall:2006yp} format, are then fed to Pythia8~\cite{Sjostrand:2014zea} for showering, hadronization, and completing the truth-level decay chains of the various produced particles.
With our recasting code (see Appendix~\ref{sec:recast}), we obtain the cutflow efficiencies $\epsilon$ for all the parameter points scanned.
Thus, we compute the signal-event numbers with the following formula,
\begin{eqnarray}
    N_S = 2\cdot  \mathcal{L}\cdot \sigma(pp\to t \bar{t})_{\text{SM}}\cdot  \mathcal{B}(t\to W^+ b) \cdot \mathcal{B}(W^+\to jj) \cdot \mathcal{B}(\bar{t}\to j a)\cdot \mathcal{B}(a\to jj) \cdot \epsilon,
\end{eqnarray}
where $\mathcal{L}$ labels the integrated luminosity, $\mathcal{B}(t\to W^+ b)=99.7\%$, and $\mathcal{B}(W^+\to jj)=67.41\%$~\cite{ParticleDataGroup:2022pth}.
Here, ``$j$'' denotes a jet including the up, down, strange, charm, and bottom quarks.
The theoretical prediction for the top-quark pair production cross section at the LHC $\sigma(pp\to t \bar{t})_{\text{SM}}=833.9$ pb is computed at NNLO+NNLL with the \texttt{Top++2.0} program, for the CM energy $\sqrt{s}=13$ TeV~\cite{Czakon:2011xx}.

\section{Numerical results}\label{sec:results}

In this section, we present and discuss the numerical results of our reinterpretation of the ATLAS DV+jets search~\cite{ATLAS:2023oti} in terms of the QFV ALP scenario. 

Reference~\cite{ATLAS:2023oti} published in its Table 6 the observed limits at 95\% C.L.~on the signal-event number, $S^{95}_{\text{obs}}=3.8$(3.0) for the High-$p_T$(Trackless)-jet SR with an integrated luminosity of 139 fb$^{-1}$.
We require these signal-event numbers for the ALP scenario in order to establish the corresponding exclusion bounds.
Further, in expectation of advancement in e.g.~technology and experimental search algorithms, we assume, for an integrated luminosity of 3000 fb$^{-1}$, the same level of background and hence the same number of expected signal-event numbers for the sensitivity reach at 95\% C.L.

\begin{table}[t]
\begin{center}
\resizebox{\textwidth}{!}{%
\begin{tabular}{l|ccc|ccc}
\hline
$m_a~[\text{GeV}], g_{ii}/\Lambda~[\text{GeV}^{-1}], c\tau_a~[\text{mm}]$         & $25, 10^{-9}, 2999$ & $25, 10^{-8}, 29.99$ & $25, 10^{-7}, 0.2999$ & $40, 10^{-9}, 1790$ & $40, 10^{-8}, 17.9$ & $40, 10^{-7}, 0.179$ \\ \hline
Jet selection                                & $9.9\times 10^{-4}$                 & $9.6\times 10^{-4}$                 & $1.0\times 10^{-3}$                 & $8.9\times 10^{-4}$                 & $8.9\times 10^{-4}$                 & $8.9\times 10^{-4}$                 \\
Event has $\geq 1$ DV passing:               &                  &                  &                  &                  &                  &                  \\
~~$R_{xy}, |z|<300$ mm                       & $1.8\times 10^{-5}$                 & $6.5\times 10^{-4}$                 & $1.0\times 10^{-3}$                 & $3.7\times 10^{-5}$                 & $8.0\times 10^{-4}$                 & $8.9\times 10^{-4}$                 \\
~~$R_{xy}>4$ mm                              & $1.7\times 10^{-5}$                 & $6.2\times 10^{-4}$                 & $1.9\times 10^{-4}$                 & $3.7\times 10^{-5}$                 & $7.5\times 10^{-4}$                 & $3.6\times 10^{-5}$                 \\
~~$\geq 1$ track with $|d_0|>2$ mm           & $1.7\times 10^{-5}$                 & $6.1\times 10^{-4}$                 & $1.5\times 10^{-4}$                 & $3.7\times 10^{-5}$                 & $7.5\times 10^{-4}$                 & $2.9\times 10^{-5}$                 \\
~~$n_{\text{selected decay products}}\geq 5$ & $1.3\times 10^{-5}$                 & $5.9\times 10^{-4}$                 & $1.4\times 10^{-4}$                 & $3.4\times 10^{-5}$                 & $7.3\times 10^{-4}$                 & $2.9\times 10^{-5}$                 \\
~~Invariant mass $> 10$ GeV                  & $7.0\times 10^{-6}$                 & $3.8\times 10^{-4}$                 & $1.1\times 10^{-4}$                 & $2.9\times 10^{-5}$                 & $6.6\times 10^{-4}$                 & $2.5\times 10^{-5}$                 \\ \hline
Param.~Effi.~                                & $2.3\times 10^{-8}$                 & $2.7\times 10^{-5}$                 & $2.3\times 10^{-5}$                 & $2.0\times 10^{-6}$                 & $1.2\times 10^{-4}$                 & $1.2\times 10^{-5}$                 \\ \hline \hline
$m_a~[\text{GeV}], g_{ii}/\Lambda~[\text{GeV}^{-1}], c\tau_a~[\text{mm}]$        & $65, 10^{-9}, 1080$ & $65, 10^{-8}, 10.8$ & $65, 10^{-7}, 0.108$ & $90, 10^{-9}, 777$ & $90, 10^{-8}, 7.77$ & $90, 10^{-7}, 0.0777$ \\ \hline
Jet selection                                & $1.0\times 10^{-3}$                 & $9.2\times 10^{-4}$                 & $9.8\times 10^{-4}$                 & $1.0\times 10^{-3}$                 & $9.5\times 10^{-4}$                 & $9.7\times 10^{-4}$                 \\
Event has $\geq 1$ DV passing:               &                  &                  &                  &                  &                  &                  \\
~~$R_{xy}, |z|<300$ mm                       & $8.4\times 10^{-5}$                 & $9.0\times 10^{-4}$                 & $9.8\times 10^{-4}$                 & $1.4\times 10^{-4}$                 & $9.4\times 10^{-4}$                 & $9.7\times 10^{-4}$                 \\
~~$R_{xy}>4$ mm                              & $8.2\times 10^{-5}$                 & $7.5\times 10^{-4}$                 & 0.0                 & $1.3\times 10^{-4}$                 & $7.3\times 10^{-4}$                 & 0.0                 \\
~~$\geq 1$ track with $|d_0|>2$ mm           & $8.1\times 10^{-5}$                 & $7.5\times 10^{-4}$                 & 0.0                 & $1.3\times 10^{-4}$                 & $7.2\times 10^{-4}$                 & 0.0                 \\
~~$n_{\text{selected decay products}}\geq 5$ & $8.0\times 10^{-5}$                 & $7.5\times 10^{-4}$                 & 0.0                 & $1.3\times 10^{-4}$                 & $7.2\times 10^{-4}$                 & 0.0                \\
~~Invariant mass $> 10$ GeV                  & $7.9\times 10^{-5}$                 & $7.2\times 10^{-4}$                & 0.0                & $1.3\times 10^{-4}$                 & $7.1\times 10^{-4}$                & 0.0                 \\ \hline
Param.~Effi.~                                & $1.3\times 10^{-5}$                 & $2.5\times 10^{-4}$                 & 0.0                & $2.8\times 10^{-5}$                 & $3.0\times 10^{-4}$                 & 0.0                \\ \hline
\end{tabular}
}
\caption{Cutflows on one million signal events with the High-$p_T$-jet search strategy for selected benchmark parameters of the ALP scenario, for $m_a=25, 40, 65,$ and 90 GeV, including the parameterized efficiencies.
The ALP's proper decay length, $c\tau_a$, is calculated with the given values of $m_a$ and $g_{ii}/\Lambda$, with $g_{ii}=g_{11}=g_{22}=g_{33}$ and with Eq.~\eqref{eq:Gamma_a2qq}.
Note that we assume the production couplings are sufficiently small so that their induced partial decay widths are negligible; in practice, we fix $g_{3i}/\Lambda=10^{-6}$ GeV$^{-1}$ for $i=1,2$ to obtain this table.
The zero entries arise because for the corresponding benchmark points the ALPs are too promptly decaying so that no event passes the selections; however, a larger data sample should result in non-zero values.
}
\label{tab:signal-cutflow-highpT}
\end{center}
\end{table}

\begin{table}[t]
\begin{center}
\resizebox{\textwidth}{!}{%
\begin{tabular}{l|ccc|ccc}
\hline
$m_a~[\text{GeV}], g_{ii}/\Lambda~[\text{GeV}^{-1}], c\tau_a~[\text{mm}]$         & $25, 10^{-9}, 2999$ & $25, 10^{-8}, 29.99$ & $25, 10^{-7}, 0.2999$ & $40, 10^{-9}, 1790$ & $40, 10^{-8}, 17.9$ & $40, 10^{-7}, 0.179$ \\ \hline
Jet selection                                & $3.1\times 10^{-3}$                 & $1.5\times 10^{-2}$                 & $1.5\times 10^{-2}$                 & $6.7\times 10^{-3}$                 & $1.5\times 10^{-2}$                 & $1.5\times 10^{-2}$                 \\
Event has $\geq 1$ DV passing:               &                  &                  &                  &                  &                  &                  \\
~~$R_{xy}, |z|<300$ mm                       & $2.3\times 10^{-4}$                 & $1.0\times 10^{-2}$                 & $1.5\times 10^{-2}$                 & $6.1\times 10^{-4}$               & $1.3\times 10^{-2}$                 & $1.5\times 10^{-2}$                 \\
~~$R_{xy}>4$ mm                              & $2.3\times 10^{-4}$                 & $9.7\times 10^{-3}$                 & $2.3\times 10^{-3}$                 & $6.0\times 10^{-4}$                 & $1.2\times 10^{-2}$                 & $2.9\times 10^{-4}$                 \\
~~$\geq 1$ track with $|d_0|>2$ mm           & $2.2\times 10^{-4}$                 & $9.6\times 10^{-3}$                 & $1.7\times 10^{-3}$                 & $6.0\times 10^{-4}$                 & $1.2\times 10^{-2}$                 & $2.3\times 10^{-4}$                 \\
~~$n_{\text{selected decay products}}\geq 5$ & $2.1\times 10^{-4}$                 & $9.2\times 10^{-3}$                 & $1.7\times 10^{-3}$                 & $5.6\times 10^{-4}$                 & $1.2\times 10^{-2}$                 & $2.3\times 10^{-4}$                 \\
~~Invariant mass $> 10$ GeV                  & $1.3\times 10^{-4}$                 & $5.9\times 10^{-3}$                & $1.2\times 10^{-3}$                 & $5.0\times 10^{-4}$                 & $1.1\times 10^{-2}$                 & $2.2\times 10^{-4}$                 \\ \hline
Param.~Effi.~                                & $6.8\times 10^{-6}$                 & $5.0\times 10^{-4}$                 & $2.4\times 10^{-4}$                 & $6.5\times 10^{-5}$                & $2.3\times 10^{-3}$                 & $7.9\times 10^{-5}$                 \\ \hline    \hline
$m_a~[\text{GeV}], g_{ii}/\Lambda~[\text{GeV}^{-1}], c\tau_a~[\text{mm}]$        & $65, 10^{-9}, 1080$ & $65, 10^{-8}, 10.8$ & $65, 10^{-7}, 0.108$ & $90, 10^{-9}, 777$ & $90, 10^{-8}, 7.77$ & $90, 10^{-7}, 0.0777$ \\ \hline
Jet selection                                & $1.3\times 10^{-2}$               & $1.7\times 10^{-2}$                 & $1.7\times 10^{-2}$                 & $1.7\times 10^{-2}$                 & $1.8\times 10^{-2}$                 & $1.8\times 10^{-2}$                 \\
Event has $\geq 1$ DV passing:               &                  &                  &                  &                  &                  &                  \\
~~$R_{xy}, |z|<300$ mm                       & $1.6\times 10^{-3}$                & $1.7\times 10^{-2}$                 & $1.7\times 10^{-2}$               & $2.9\times 10^{-3}$                 & $1.8\times 10^{-2}$                 & $1.8\times 10^{-2}$                 \\
~~$R_{xy}>4$ mm                              & $1.6\times 10^{-3}$                 & $1.4\times 10^{-2}$                & $4.0\times 10^{-6}$                 & $2.9\times 10^{-3}$                 & $1.3\times 10^{-2}$                 & 0.0                 \\
~~$\geq 1$ track with $|d_0|>2$ mm           & $1.6\times 10^{-3}$                 & $1.4\times 10^{-2}$                 & $3.0\times 10^{-6}$                 & $2.9\times 10^{-3}$                 & $1.3\times 10^{-2}$                 & 0.0                \\
~~$n_{\text{selected decay products}}\geq 5$ & $1.6\times 10^{-3}$                 & $1.4\times 10^{-2}$                 & $3.0\times 10^{-6}$              & $2.9\times 10^{-3}$                 & $1.3\times 10^{-2}$                 & 0.0                 \\
~~Invariant mass $> 10$ GeV                  & $1.5\times 10^{-3}$                 & $1.3\times 10^{-2}$                 & $3.0\times 10^{-6}$                 & $2.8\times 10^{-3}$                 & $1.3\times 10^{-2}$                 & 0.0                \\ \hline
Param.~Effi.~                                & $2.7\times 10^{-4}$                 & $5.4\times 10^{-3}$                 & $1.2\times 10^{-6}$                 & $6.2\times 10^{-4}$                & $6.8\times 10^{-3}$                 & 0.0                 \\ \hline
\end{tabular}
}
\caption{The same table as Table~\ref{tab:signal-cutflow-highpT}, but for the Trackless-jet search strategy.
}
\label{tab:signal-cutflow-trackless}
\end{center}
\end{table}

We first list in Table~\ref{tab:signal-cutflow-highpT} and Table~\ref{tab:signal-cutflow-trackless} the cutflows including the acceptances and the final parameterized efficiencies on one million signal events for selected example benchmark parameters of our signal model, for the High-$p_T$-jet SR and Trackless-jet SR, respectively.
For $m_a=25, 40, 65,$ and 90 GeV, we consider the decay couplings $g_{ii}/\Lambda=10^{-9}, 10^{-8},$ and $10^{-7}$ GeV$^{-1}$ and compute the corresponding proper decay length of the ALP, $c\tau_a$, collectively listed in the tables.
In particular, for simplicity, we assume negligibly small production couplings so that their induced decays of the ALP can be safely neglected; specifically, we set $g_{3i}/\Lambda=10^{-6}$ GeV$^{-1}$ with $i=1,2$ for deriving Table~\ref{tab:signal-cutflow-highpT} and Table~\ref{tab:signal-cutflow-trackless}.
We also note that, here, $g_{ii}$ denotes the decay couplings $g_{11}=g_{22}=g_{33}$.

\begin{figure}[t]
	\centering
	\includegraphics[width=0.495\textwidth]{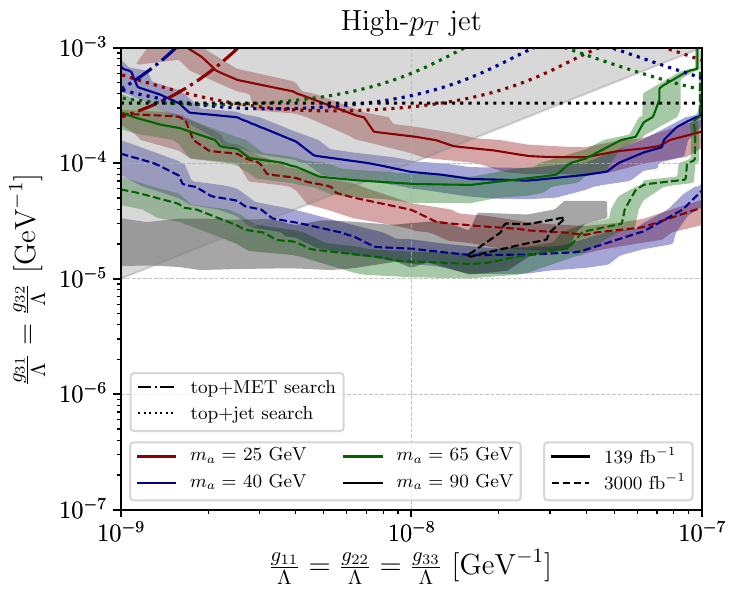}
	\includegraphics[width=0.495\textwidth]{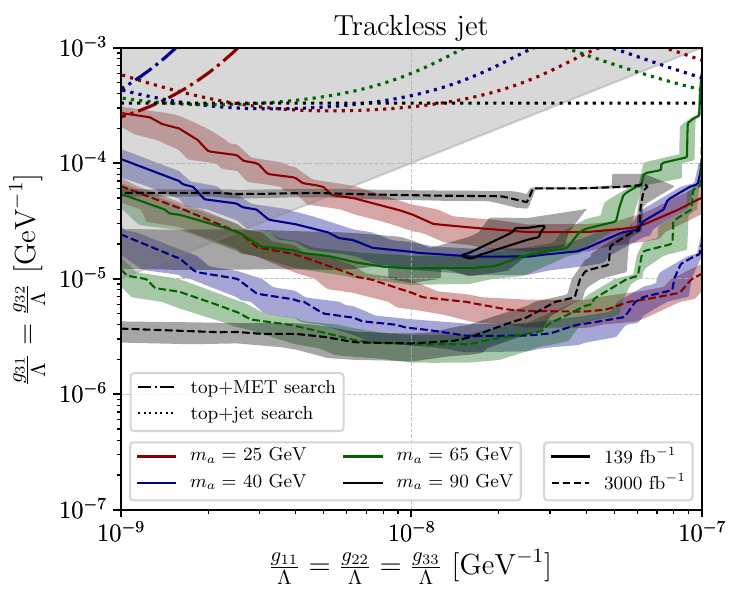}
	\caption{ATLAS sensitivity reach at $95\%$ C.L.~with 139 fb$^{-1}$ (solid) and 3 ab$^{-1}$ (dashed) integrated luminosities, in the plane $\frac{g_{31}}{\Lambda}=\frac{g_{32}}{\Lambda}$ vs.~$\frac{g_{11}}{\Lambda}=\frac{g_{22}}{\Lambda}=\frac{g_{33}}{\Lambda}$, for various ALP mass choices, with the ``High-$p_T$ jet'' (left) and ``Trackless jet'' (right) search strategies.
    An error band at 50\% level is displayed together.
    The gray hatched region is where our results do not apply because we do not include the 4-body decay modes of the ALP induced by the production couplings for ALP masses below roughly 85 GeV.
    Bounds obtained in Ref.~\cite{Esser:2023fdo} by recasting the ATLAS search~\cite{ATLAS:2021hza} is weak and outside the parameter range displayed here, and is hence not shown.
    The dotted and the dot-dashed curves correspond to reinterpretation results of a CMS~\cite{CMS:2016uzc} and an ATLAS~\cite{ATLAS:2015iqc} search, taking into account the upper bounds on potential new-physics contributions to the cross section of $pp\to t j$ and $pp\to t+$ MET, respectively; see Appendix~\ref{sec:recast_single_top} for the detail.
 }
 \label{fig:sensitivities_g_vs_g}
\end{figure}

Now we present in Fig.~\ref{fig:sensitivities_g_vs_g} sensitivity reach of the ATLAS DV+jets search in the parameter plane $\frac{g_{31}}{\Lambda}=\frac{g_{32}}{\Lambda}$ vs.~$\frac{g_{11}}{\Lambda}=\frac{g_{22}}{\Lambda}=\frac{g_{33}}{\Lambda}$, with the ALP mass fixed at 25, 40, 65, and 90 GeV.

The left(right) panel corresponds to the numerical results obtained with the ``High $p_T$''(``Trackless'')-jet search SR.
The solid and dashed lines are for an integrated luminosity of 139 fb$^{-1}$ and 3000 fb$^{-1}$, respectively.
The regions above these lines are excluded by the search, except for the case of $m_a=90$ GeV for which the area inside the bounded regions is excluded.
Furthermore, the dotted and dot-dashed lines are derived by recasting a CMS search~\cite{CMS:2016uzc} for a single top plus jets and an ATLAS search~\cite{ATLAS:2015iqc} for a single top with FCNC interactions, respectively.
We leave the detail in Appendix~\ref{sec:recast_single_top}.
We apply a conservative $50\%$ error band on our results taking into account the uncertainties in the cutflow efficiencies, in particular, considering the non-closure we observe in our recast validation which can be up to about 75\% in some benchmarks (see Appendix~\ref{sec:recast}), as well as the uncertainties in our computation of the ALP's decay widths.
Note that the gray hatched region is where the production couplings are at least 4 orders of magnitude larger than the decay couplings, and thus, the 4-body decay widths of the ALPs with a mass below about 85 GeV induced by the production couplings are important or even dominant.
Since we ignore these modes for reasons of computational resources, this gray hatched region is where our sensitivity reach is less accurate (see the discussion at the end of Sec.~\ref{sec:model}).

In general, we find that the two SRs are sensitive to similar decay-coupling ranges and hence similar lifetimes of the ALP.
For the sensitivity reach in the production couplings, however, we observe that the Trackless-jet SR performs much better than the High-$p_T$-jet SR by a factor of $\sim 5$.
For the HL-LHC integrated luminosity of 3000 fb$^{-1}$ the sensitivity reach in the production couplings is stronger than that for the LHC Run 2 integrated luminosity of 139 fb$^{-1}$ by roughly $\sim 5$; this is because the signal-event number is proportional to the production couplings squared and thus re-scaling the integrated luminosities gives $\sqrt{3000/139}\sim 4.6$.

The existing single-top searches at the LHC can compete with only the High-$p_T$-jet SR in the DV+jets search.
We find that the search for top plus prompt jets can be complementary to the High-$p_T$-jet SR for the cases of $m_a=25, 40,$ and 65 GeV in small and large decay coupling regions as shown in the left plot of Fig.~\ref{fig:sensitivities_g_vs_g}.
For $m_a=90$ GeV, the opened channel of the ALP's 3-body decays via the relatively large production couplings saturates the ALP decay widths, and, thus, the reinterpreted bounds are a horizontal line, implying independence of the decay couplings $\frac{g_{ii}}{\Lambda}$.
We observe these bounds are rather weak compared to those derived in the High-$p_T$-jet SR.

Both SRs show sensitivities to all the benchmark ALP masses studied, though the High-$p_T$-jet SR has no sensitivity to the case of $m_a=90$ GeV with 139 fb$^{-1}$ integrated luminosity. 
For $m_a=90$ GeV, the production couplings considered can lead to 3-body decays of the ALP into an up or charm quark, a $W$-boson, and a $b$-quark.
Thus, the sensitivity reaches for $m_a=90$ GeV are bounded from above as too large values of the production couplings would render the ALP too promptly decaying.
Moreover, we observe that for $m_a=90$ GeV, some sensitivity curves and all the error bands extend all the way to the very left end horizontally.
This is because in these parts of the parameter space, both the production and the decay of the ALP are dominated by the couplings $g_{3i}/\Lambda$, and the decay couplings $g_{ii}/\Lambda$ make only negligible contributions to the ALP's total decay width.
In contrast, we find that for the High-$p_T$-jet(Trackless-jet) SR, the 3000 fb$^{-1}$(139 fb$^{-1}$) sensitivity curve of the corresponding $S^{95}_{\text{obs}}$ signal events shows no sensitivities to smaller values of the decay couplings $g_{ii}/\Lambda$.
This is because in these parameter regions, the production couplings alone do not lead to sufficiently many signal events required, and hence relatively large decay couplings are required in order to enhance the fiducial-volume selection efficiencies so as to reach $N_S=S^{95}_{\text{obs}}$.

\begin{figure}[t]
	\centering
	\includegraphics[width=0.495\textwidth]{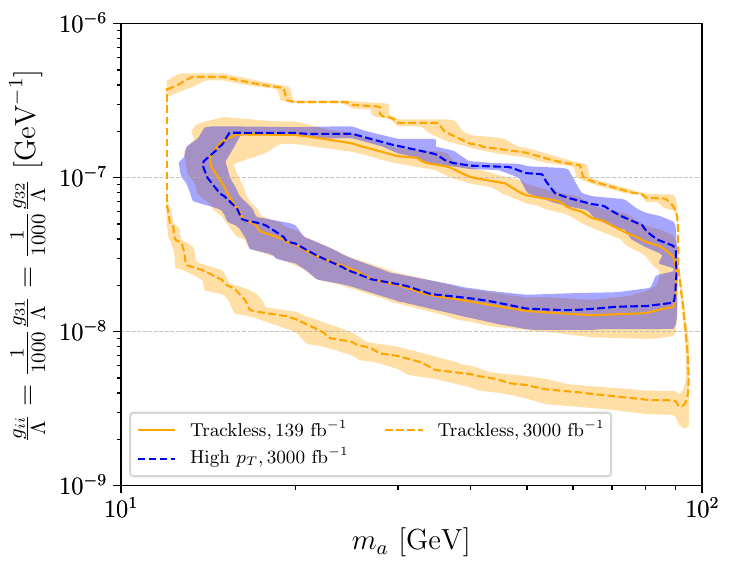}
	\includegraphics[width=0.495\textwidth]{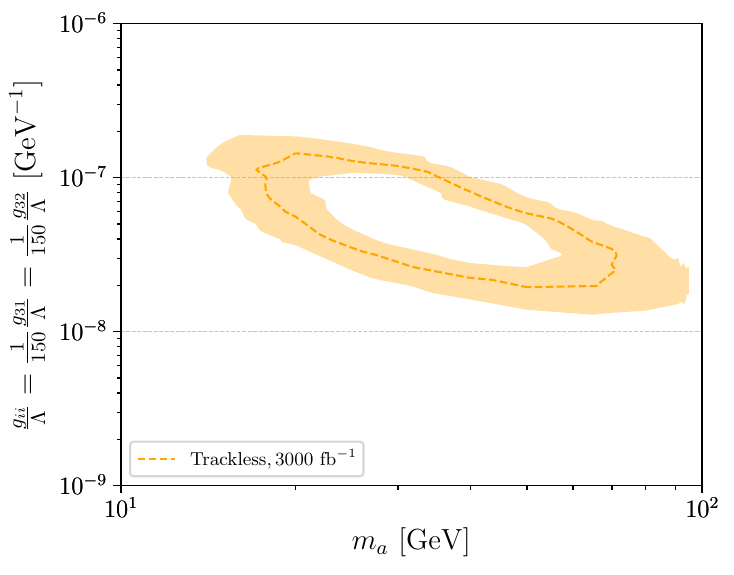}
	\caption{ATLAS sensitivity reach at $95\%$ C.L.~with 139 fb$^{-1}$ and 3 ab$^{-1}$ integrated luminosities, respectively, in the plane $\frac{g_{ii}}{\Lambda}=\frac{1}{x}\frac{g_{31}}{\Lambda}=\frac{1}{x}\frac{g_{32}}{\Lambda}$ vs.~$m_a$ for $x=1000$ (left panel) and $x=150$ (right panel), with the two search strategies.
    $g_{ii}$ labels the universal quark-flavor-diagonal couplings with $g_{ii}=g_{11}=g_{22}=g_{33}$.
    As in Fig.~\ref{fig:sensitivities_g_vs_g}, error bands at the level of 50\% are shown.
 }
 \label{fig:sensitivities_g_vs_m}
\end{figure}

Moving to Fig.~\ref{fig:sensitivities_g_vs_m}, we show two plots in the plane $\frac{g_{ii}}{\Lambda}=\frac{1}{x}\frac{g_{31}}{\Lambda}=\frac{1}{x}\frac{g_{32}}{\Lambda}$ vs.~$m_a$, for $x=1000$ and 150, respectively.
In the left plot where the production couplings are assumed to be 1000 times the decay couplings, the High-$p_T$-jet SR is found to be sensitive only for $\mathcal{L}=3000$ fb$^{-1}$, while the Trackless-jet SR can probe the model for both 139 fb$^{-1}$ and 3000 fb$^{-1}$ integrated luminosities.
We observe that for the ALP mass between about 12 GeV and 95 GeV, the Trackless-jet SR can probe the decay couplings $g_{ii}/\Lambda$, depending on the ALP mass, between about $3\times 10^{-9}$ GeV$^{-1}$ and $5\times 10^{-7}$ GeV$^{-1}$, for $\mathcal{L}=3000$ fb$^{-1}$.
In the right plot, we assume smaller production couplings which are now 150 times the decay ones.
This implies smaller production rates of the long-lived ALP, and, as a result, only the most sensitive SR, the Trackless-jet SR, with an integrated luminosity of 3000 fb$^{-1}$, would be sensitive to certain relatively small parts of the parameter space.
Similar to Fig.~\ref{fig:sensitivities_g_vs_g}, we assume an uncertainty level at 50\% and show the corresponding error bands in these plots.
For the considered fixed relations between the production and decay couplings, we find that the single top plus jets or MET searches cannot compete with the ATLAS DV+jets search for the whole ALP mass range, and their exclusion bounds are hence not shown.

The sensitive regions are bounded from above; otherwise the decay widths would be too large for the ALP to decay inside the considered fiducial volume, which we define with some of the vertex-level acceptance selections: $R_{xy}, |z| < 300$ mm and $R_{xy}>4$ mm.
For the same reason, the sensitivity reach has an upper mass bound just below 100 GeV.
The lower mass reach is mainly due to the invariant-mass requirement $m_{\text{DV}}>10$ GeV in the vertex selections of the ATLAS search.
Below the lower bounds of the sensitivity reach to the ALP couplings, the ALP production rates would be so small and the ALP decay length would be so long that the ALP tends to decay only after passing through the ATLAS detector's fiducial volume, and these two effects combined lead to insufficient signal-event rates.

We further explain the overall slope of the ellipse-like sensitivity shape.
This is mainly due to the effect of the ALP mass on the total decay width of the ALP, as generally speaking increasing the mass also enhances the total decay width.
In the large-coupling regime, the ALP is rather prompt; as a result, with a larger $m_a$, the ALP becomes more prompt leading to even smaller decay probabilities of the ALP inside the fiducial volume.
On the other hand, in the small-coupling limit, the ALP is already long-lived, where the signal-event rate is actually proportional to the ALP's total decay width.
This can be understood in the following way.
The signal-event number is proportional to the decay probability of the ALP inside the fiducial volume, which can be computed with the exponential decay law, and the exponential functions can be expanded if the boosted decay length is much larger than the detector's distance to the IP; as a result, the signal-event number would then be proportional to the ALP total decay width in the large decay-length, or equivalently, the small ALP-couplings limit.
Therefore, as long as we are in this limit, a heavier ALP implies a larger decay width which then enhances the signal-event rates.

We note that the parameterized efficiencies provided by the ATLAS collaboration -- while designed to be as model-independent as possible -- were validated only for the RPV benchmarks tested by the collaboration, and therefore should be applied with caution for other models. 
We thus use these parameterized efficiencies while also discussing and checking the relevance of the scope of the efficiencies to our ALP scenarios before ending this section.
The recasting instruction lists three conditions, under which the parameterized efficiencies are validated and can be used safely.
Briefly summarized, these conditions require that:
\begin{enumerate}
    \item the event- and vertex-level acceptances combined should exceed $10\%$;
    \item for any parameter-space point, if the corresponding signal events have an acceptance of over 90\% with the High-$p_T$-jet SR, the parameterized efficiencies of the Trackless-jet SR should not be used; and
    \item the LLP proper decay length of less than 3 meters are recommended for models where ``jets primarily originate from the decay of LLPs''.
\end{enumerate}
We examine our signal-event cutflows (see e.g.~Table~\ref{tab:signal-cutflow-highpT} and Table~\ref{tab:signal-cutflow-trackless}) closely against these conditions.
Apparently, Condition 1 is not met in most, if not all, of the parameter-space points of the ALP scenario that we consider for both SRs.
Consequently, the signal-event acceptances of the High-$p_T$-jet SR are all less than 90\% and therefore, Condition 2 is automatically well satisfied, so that the parameterized efficiencies of the trackless-jet SR can be used safely.
As for Condition 3, in our ALP model, jets stem not only from the ALP decays, but also, importantly, from the top quarks' decays; as a result, this condition is not closely relevant for our study.
In summary, we find our ALP scenario does not satisfy Condition 1 but Condition 2 is fulfilled, while the final condition is irrelevant.
Nevertheless, this does not mean that the parameterized efficiencies are forbidden from being used in the ALP scenario and it only implies that such a reinterpretation should be regarded with caution and further validation from the experimentalists should be performed.

\section{Summary and outlook}\label{sec:conclusions}

Recently, ATLAS published a search~\cite{ATLAS:2023oti} for DV and multiple jets, obtaining latest bounds on long-lived electroweakinos in the RPV-SUSY which decay via baryon-number-violating operators $\lambda'' \bar{U} \bar{D} \bar{D}$.
The search has two SRs, High-$p_T$ jet and Trackless jet, and both strong and electroweak production channels of the electroweakinos are considered.
In this work, we have recast this search and validated our code by comparing cutflow efficiencies.
We achieve excellent agreement at the acceptance level.
However, once the parameterized efficiencies, provided by the ATLAS collaboration to account for further selections such as multi-jet trigger and detector material map, are included, the validation performance is not the optimal.
In particular, we find good agreement only in some benchmarks while in other benchmarks our results can be off from the experimental full simulation by up to about 75\%.

We then proceed to apply our recasting code in a theoretical scenario with an ALP coupled both diagonally and off-diagonally to quarks.
Specifically, we focus on QFV couplings of the ALP with the top quark and the up/charm quark, primarily mediating the production of the ALP from top-quark decays.
The diagonal couplings mainly lead to the tree-level decays of the ALP into jets.
The ALP can be long-lived for sufficiently small couplings to the quarks, thus potentially forming a DV inside the fiducial volume (tracker) of the ATLAS detector.
With our recasting code, we obtain the cutflow efficiencies at each scanned parameter point of the model, and then compute the expected signal-event numbers with either the current 139 fb$^{-1}$ or the future HL-LHC 3000 fb$^{-1}$ integrated luminosity.
This allows for obtaining the exclusion bounds at 95\% C.L.
We show our numerical results of the sensitivity reach in two figures.
First, we fix the ALP mass at some representative values, and present the bounds in the plane of the production couplings vs.~the decay couplings.
We find both SRs are sensitive to certain parts of the parameter points, and the Trackless-jet SR performs better than the High-$p_T$-jet SR.
The main reason is that for the ALP mass range studied it is easier to pass the jet-selection criteria of the Trackless-jet SR than those of the High-$p_T$-jet SR.
In particular, for 139 fb$^{-1}$, the Trackless-jet SR can probe the ALP parameter space for the ALP mass even above the threshold of $m_W+m_b\sim 85$ GeV, touching a unique part of the parameter space.
We also find that existing single-top plus jets or MET searches at the LHC can be complementary to the ATLAS DV+jets search.
We then fix the proportionality relations between the production and the decay couplings, and present sensitivity reach in the plane of the decay couplings vs.~the ALP mass.
We find the search with either 139 fb$^{-1}$ or 3000 fb$^{-1}$ integrated luminosity can be sensitive to the decay couplings $\frac{g_{ii}}{\Lambda}$ of order $\mathcal{O}(10^{-9})-\mathcal{O}(10^{-7})$ GeV$^{-1}$, for the ALP mass roughly between 12 GeV and 95 GeV and the production couplings being 1000 times the decay ones.
For another benchmark where the production couplings are 150 times the decay ones, weaker sensitivities are found.
In addition, in Table~\ref{tab:signal-cutflow-highpT} and Table~\ref{tab:signal-cutflow-trackless} we list cutflows for representative mass and decay couplings of the ALP.

We conclude that the ATLAS search can probe unique parts of the parameter space of the QFV ALP scenario considered here, which are currently not excluded.
This sheds light on potential further applications of the search in other theoretical scenarios predicting signatures of displaced vertices plus jets.

\section*{Acknowledgment}

We thank David Rousso, Emily Thompson, and Catherine Xu, for useful discussions concerning the recast of the ATLAS DV+jets search.
We thank Jong Soo Kim and Kechen Wang for help with the jet merging procedure.
G.C.~acknowledges support from ANID FONDECYT grant No.~11220237 and ANID – Millennium Science Initiative Program ICN2019\_044.
The work was supported by the MoST of Taiwan under Grants MOST-110-2112-M-007-017-MY3.

\appendix
\section{Recast of the DV+jets search}\label{sec:recast}

With MadGraph5 3.4.1~\cite{Alwall:2011uj,Alwall:2014hca}, we generate signal-event sample LHE files in the theoretical framework of the MSSM.
The SUSY SLHA spectrum files are provided by the ATLAS collaboration, where we only need to tune the gluino mass, electroweakinos' masses, as well as the total decay width of the electroweakinos according to the selected benchmarks given in Ref.~\cite{ATLAS:2023oti}.
All the other SUSY particles' masses have been set to very large decoupled values.
For the strong-production channel, we simulate $pp\to \tilde{g}\tilde{g}$ events at the CM energy $\sqrt{s}=13$ TeV, where the gluinos decay promptly to the lightest neutralino (which is mainly bino-like) and two light jets with equal branching ratios shared between  the down, up, strange, and charm quarks: $\tilde{g}\to \tilde{\chi}^0_1 j j$, and the lightest neutralino is set to decay to three light quarks via $\lambda'' \bar{U}\bar{D}\bar{D}$ operators.
For the EW-production channels, pair-production processes of the mass-degenerate electroweakinos $\tilde{\chi}^0_1, \tilde{\chi}^0_2,$ and $\tilde{\chi}^\pm_1$ are taken into account: $pp\to \tilde{\chi}^0_1 \tilde{\chi}^+_1, \tilde{\chi}^0_1 \tilde{\chi}^-_1, \tilde{\chi}^-_1 \tilde{\chi}^+_1, \tilde{\chi}^0_1  \tilde{\chi}^0_2, \tilde{\chi}^0_2 \tilde{\chi}^+_1, \tilde{\chi}^0_2 \tilde{\chi}^-_1$.
The electroweakinos, which are assumed to be pure Higgsinos, are then set to decay to either light or heavy quarks according to the benchmarks studied on hand.

The hard-process events are simulated together with up to two additional jets.
We perform the CKKW-L jet merging scheme~\cite{Lonnblad:2001iq}, setting in the \texttt{run\_card.dat} of MadGraph5 that the value of \texttt{ktdurham} should be equal to the jet merging scale, which is a quarter of the gluino mass in the strong-channel processes and a quarter of the electroweakino masses in the EW processes.
The generated LHE files are then showered and hadronized in Pythia8.308~\cite{Sjostrand:2014zea} where we switch on \texttt{kT} merging with the jet-merging scale set to a quarter of the SUSY particles produced (equal to the value of \texttt{ktdurham} set in the \texttt{run\_card.dat} of MadGraph5).
It is important that we switch on \texttt{Merging:mayRemoveDecayProducts} in order to perform merging before the decay products of the resonances are generated in Pythia.

Note that during the event generation, \texttt{NNPDF2.3lo}~\cite{
Ball:2012cx} is chosen for the parton distribution function of the protons, with the \texttt{A14}~\cite{ATL-PHYS-PUB-2014-021} tune.

\textit{Pre-selections}: the default parton-level selections in MadGraph5 are used.
We implement a toy-detector module in Pythia for reconstructing truth-level jets, in a similar way as in Ref.~\cite{Allanach:2016pam} but matching the truth-level selection for jets described in the HEPData recast instruction note for the DV+jets ATLAS search~\cite{ATLAS:2023oti}.

Truth jets are reconstructed with FastJet~\cite{Cacciari:2011ma}, using anti-$k_t$ algorithm and $R=0.4$ from all selected stable particles excluding neutrinos and muons.
This definition includes particles from the LLP decay.
Jet momentum smearing is applied with formulas given in Ref.~\cite{Allanach:2016pam}.
Further, displaced jets are defined as those among the jets selected above that are matched with the LLPs' decay positions and have $|\eta|<2.5$.
By calculating $\Delta R$ between the LLP decay products and the truth jets, we determine that a truth jet stems from the LLP decay if the closest decay products of the LLP has $\Delta R<0.3$.
Furthermore, for the displaced truth jets, we require that the matched LLPs should decay with a transverse distance from IP smaller than 3870 mm which corresponds to the region of the calorimeter.

\begin{table}[]
\begin{center}
\begin{tabular}{c|c|c}
SR            & High-$p_T$ jet                                                  & Trackless jet                                                                  \\ \hline
              & $n^{250}_{\text{jet}}\geq 4$ or $n^{195}_{\text{jet}}\geq 5$    & $n^{137}_{\text{jet}}\geq 4$ or $n^{101}_{\text{jet}}\geq 5$                   \\
Jet selection & or $n^{116}_{\text{jet}}\geq 6$  or $n^{90}_{\text{jet}}\geq 7$ & or $n^{83}_{\text{jet}}\geq 6$ or $n^{55}_{\text{jet}}\geq 7$,                    \\
              &                                                                 & $n^{70}_{\text{displaced jet}}\geq 1$ or $n^{50}_{\text{displaced jet}}\geq 2$ \\ \hline
\end{tabular}
\caption{Truth-jet selection requirements. $n^{250}_{\text{jet}}\geq 4$ means at least 4 jets should have a $p_T$ larger than or equal to 250 GeV, and similarly for the other notations.}
\label{tab:jet_pt_requirement}
\end{center}
\end{table}

\textit{Event selections}: we follow step by step the recasting instruction provided on HEPData by the ATLAS collaboration for this search.
Event-level requirement of certain numbers of jets with different $p_T$ thresholds is first applied onto the events.
The detailed selections differ between the High-$p_T$-jet and the Trackless-jet SRs.
We reproduce these requirements from Ref.~\cite{ATLAS:2023oti} here in Table~\ref{tab:jet_pt_requirement}.
For the events passing the jet selections, at least one vertex should pass a list of vertex requirements, in order to obtain the vertex-level acceptance: 
\begin{enumerate}
    \item Both the transverse distance $R_{xy}$ and the absolute longitudinal distance $|z|$ of the vertex from the IP should be smaller than 300 mm.
    \item $R_{xy}$ should further be larger than 4 mm.
    \item At least one track should have a transverse impact parameter satisfying $d_0>2$ mm.
    \item The displaced vertex should have at least 5 decay products of a massive particle satisfying the following requirements (``selected decay products''):
    \begin{enumerate}
        \item It should be a track with a boosted transverse decay length larger than 520 mm.
        \item Its $p_T$ and charge $q$ should obey the relation $p_T/|q|> 1$ GeV.
    \end{enumerate}
    \item The truth vertex should have an invariant mass larger than 10 GeV, which is constructed with the decay products passing the above requirements, for which the mass of each decay product is assumed to be that of a charged pion.
\end{enumerate}

For events that have passed the above event and vertex acceptance requirements, we make use of parameterized efficiencies provided by the ATLAS collaboration at both event and vertex levels that account for quality requirements such as multi-jet trigger and material veto that are difficult to simulate.
The event-level efficiencies $\epsilon_{\text{event}}$ are functions of the truth-jet scalar $p_T$ sum and the transverse distance of the furthest LLP decay.
The vertex-level efficiencies $\epsilon_{\text{vertex}}$ are for reconstructing the DVs, and are functions of the DV's transverse distance to the IP, its invariant mass, as well as the LLP decay product multiplicity.

We compute the final cutflow efficiency with the following formula,
\begin{eqnarray}
    \epsilon = \mathcal{A}_{\text{event}} \cdot \epsilon_{\text{event}} \cdot \Big( 1 - {\displaystyle \prod_{\text{vertex}}} (1 - \mathcal{A}_{\text{vertex}} \cdot \epsilon_{\text{vertex}}) \Big),
\end{eqnarray}
where $\mathcal{A}_{\text{event}}$ and $\mathcal{A}_{\text{vertex}}$ label the portion of events satisfying the event-level and vertex-level acceptance requirements, respectively.

In the following, we show tables listing the cutflow efficiency results' comparison, considering both strong-channel and EW-channel benchmarks given in the recasting instruction; this includes not only light-flavor cases where the long-lived electroweakinos decay to light quarks, but also heavy-flavor ones.
The heavy-flavor benchmarks are included for recasting validation, because in the ALP scenario we study, the long-lived ALP also decays to $b$-quarks.

\begin{table}[t]
\begin{center}
\resizebox{\textwidth}{!}{%
\begin{tabular}{|l|cccccccc}
\hline
                                           & \multicolumn{8}{|c|}{Acceptance $[\%]$}                                                                                                                                                                               \\ \cline{2-9} 
 \multicolumn{1}{|c|}{$m(\tilde{g})$ [GeV]}                   & \multicolumn{2}{c|}{2000}                             & \multicolumn{2}{c|}{2000}                             & \multicolumn{2}{c|}{2400}                             & \multicolumn{2}{c|}{2000}                    \\
 \multicolumn{1}{|c|}{$m(\tilde{\chi}^0_1)$ [GeV]}               & \multicolumn{2}{c|}{850}                              & \multicolumn{2}{c|}{50}                               & \multicolumn{2}{c|}{200}                              & \multicolumn{2}{c|}{1250}                    \\
 \multicolumn{1}{|c|}{$\tau(\tilde{\chi}^0_1)$ [ns]  }            & \multicolumn{2}{c|}{0.01}                             & \multicolumn{2}{c|}{0.1}                              & \multicolumn{2}{c|}{1}                                & \multicolumn{2}{c|}{10}                      \\ \hline
Selection                                  & Exp.                 & \multicolumn{1}{c|}{This work} & Exp.                 & \multicolumn{1}{c|}{This work} & Exp.                 & \multicolumn{1}{c|}{This work} & Exp.                 & \multicolumn{1}{c|}{This work  }           \\
Jet selection                              & 99.9                 & \multicolumn{1}{c|}{99.8}          & 96.6                 & \multicolumn{1}{c|}{96.9}          & 97.2                 & \multicolumn{1}{c|}{98.2}          & 96.1                 & \multicolumn{1}{c|}{99.9} \\
Event has $\geq 1$ DV passing:             & \multicolumn{1}{l}{} & \multicolumn{1}{l|}{}          & \multicolumn{1}{l}{} & \multicolumn{1}{l|}{}          & \multicolumn{1}{l}{} & \multicolumn{1}{l|}{}          & \multicolumn{1}{l}{} & \multicolumn{1}{c|}{} \\
~~$R_{xy}, |z|<300$ mm                     & 99.9                 & \multicolumn{1}{c|}{99.8}          & 78.7                 & \multicolumn{1}{c|}{79.7}          & 44.7                 & \multicolumn{1}{c|}{45.5}          & 31.7                 &        \multicolumn{1}{c|}{31.2}               \\
~~$R_{xy}>4$ mm                            & 29.6                 & \multicolumn{1}{c|}{29.7}          & 77.0                 & \multicolumn{1}{c|}{78.3}          & 43.8                 & \multicolumn{1}{c|}{44.7}          & 30.9                 &        \multicolumn{1}{c|}{30.5}               \\
~~$\geq 1$ track with $|d_0|>2$ mm         & 29.6                 & \multicolumn{1}{c|}{29.7}          & 75.6                 & \multicolumn{1}{c|}{77.6}          & 43.7                 & \multicolumn{1}{c|}{44.7}          & 30.9                 &       \multicolumn{1}{c|}{30.5}                \\
~~$n_{\text{selected decay products}}\geq 5$ & 29.6                 & \multicolumn{1}{c|}{29.7}          & 75.5                 & \multicolumn{1}{c|}{77.3}          & 43.7                 & \multicolumn{1}{c|}{44.7}          & 30.9                 &       \multicolumn{1}{c|}{30.5}                \\
~~Invariant mass $> 10$ GeV                  & 29.6                 & \multicolumn{1}{c|}{29.7}          & 74.7                 & \multicolumn{1}{c|}{75.8}          & 43.7                 & \multicolumn{1}{c|}{44.7}          & 30.9                 &      \multicolumn{1}{c|}{30.5}            \\ 
\hline     
\end{tabular}
}
\caption{High-$p_T$-jet SR acceptance with full cutflow. With the strong-channel production and the lightest neutralino decaying to light-flavor quarks.}
\label{tab:highpt_LF_acc_full}
\end{center}
\end{table}

In Table~\ref{tab:highpt_LF_acc_full}, we compare the full cutflow with acceptances only, for the High-$p_T$-jet SR in the strong-channel production of the lightest neutralinos which decay to light-flavor quarks.
We find in all four benchmarks, the validation works very well in each step of the event-level and vertex-level acceptance selections.

\begin{table}[t]
\begin{center}
\begin{tabular}{c| c c c c}
\hline
$m(\tilde{g}), m(\tilde{\chi}^{0}_{1}), \tau(\tilde{\chi}^{0}_{1})$ & Full Sim. & Param.~Exp. & Param.~Ours & Non-closure \\
\hline
2000 GeV, 850 GeV, 0.01 ns & 27.8$\%$ & 26.0$\%$ & 26.6$\%$ & -4.3$\%$ \\
2000 GeV, 50 GeV, 0.1 ns & 14.4$\%$ & 13.8$\%$ & 21.6$\%$ & 50.0$\%$ \\
2400 GeV, 200 GeV, 1 ns & 11.5$\%$ & 11.5$\%$ & 14.4$\%$ & 25.2$\%$ \\
2000 GeV, 1250 GeV, 10 ns & 9.2$\%$ & 8.6$\%$ & 8.4$\%$ & -8.7$\%$ \\
\hline
\end{tabular}
\caption{High-$p_T$-jet SR $\epsilon$ including both the acceptances' and the parameterized efficiencies' effects. With the strong-channel production and the lightest neutralino decaying to light-flavor quarks.}
\label{tab:highpt_LF_acc_eff}
\end{center}
\end{table}

In Table~\ref{tab:highpt_LF_acc_eff}, we show results where now the parameterized efficiencies have been included in the computation.
We compare the results of the experimental full simulation, recast results with the parameterized efficiencies given by the ATLAS collaboration, as well as our own recast results with the parameterized efficiencies.
The final column ``Non-closure'' is defined as the difference between the full-simulation results and our recast results with respect to the full-simulation results, given in percentage.
We observe that our recast is validated well in the first and the fourth benchmarks, while for the second and third benchmarks, non-closure is about 50\% and 30\%, respectively.

\begin{table}[t]
\begin{center}
\resizebox{\textwidth}{!}{%
\begin{tabular}{|l|cccccccc}
\hline
                                           & \multicolumn{8}{c|}{Acceptance $[\%]$}                                                                                                                                                                               \\ \cline{2-9} 
\multicolumn{1}{|c|}{$m(\tilde{\chi}^0_1)$ [GeV]}                & \multicolumn{2}{c|}{500}                              & \multicolumn{2}{c|}{500}                               & \multicolumn{2}{c|}{1300}                              & \multicolumn{2}{c|}{1300}                    \\
\multicolumn{1}{|c|}{$\tau(\tilde{\chi}^0_1)$ [ns]}              & \multicolumn{2}{c|}{0.1}                             & \multicolumn{2}{c|}{1}                              & \multicolumn{2}{c|}{0.1}                                & \multicolumn{2}{c|}{1}                      \\ \hline
Selection                                  & Exp.                 & \multicolumn{1}{c|}{This work} & Exp.                 & \multicolumn{1}{c|}{This work} & Exp.                 & \multicolumn{1}{c|}{This work} & Exp.                 & \multicolumn{1}{c|}{This work}           \\
Jet selection                              & 49.5                 & \multicolumn{1}{c|}{51.0}          & 50.1                 & \multicolumn{1}{c|}{51.0}          & 96.8                 & \multicolumn{1}{c|}{98.5}          & 98.5                & \multicolumn{1}{c|}{98.5} \\
Event has $\geq 1$ DV passing:             & \multicolumn{1}{l}{} & \multicolumn{1}{l|}{}          & \multicolumn{1}{l}{} & \multicolumn{1}{l|}{}          & \multicolumn{1}{l}{} & \multicolumn{1}{l|}{}          & \multicolumn{1}{l}{} & \multicolumn{1}{c|}{} \\
~~$R_{xy}, |z|<300$ mm                     & 49.5                 & \multicolumn{1}{c|}{51.0}          & 41.0                 & \multicolumn{1}{c|}{41.5}          & 96.8                 & \multicolumn{1}{c|}{98.5}          & 92.1                 &             \multicolumn{1}{c|}{92.4}         \\
~~$R_{xy}>4$ mm                            & 46.5                & \multicolumn{1}{c|}{47.6}          & 39.8                & \multicolumn{1}{c|}{40.4}          & 85.9                & \multicolumn{1}{c|}{86.9}          & 89.9                 &           \multicolumn{1}{c|}{90.5}           \\
~~$\geq 1$ track with $|d_0|>2$ mm         & 46.5                 & \multicolumn{1}{c|}{47.6}          & 39.8                 & \multicolumn{1}{c|}{40.4}          & 85.9                 & \multicolumn{1}{c|}{86.9}          & 89.9                 &            \multicolumn{1}{c|}{90.5}          \\
~~$n_{\text{selected decay products}}\geq 5$ & 46.5                 & \multicolumn{1}{c|}{47.6}          & 39.8                & \multicolumn{1}{c|}{40.4}          & 85.9                 & \multicolumn{1}{c|}{86.9}          & 89.9                &          \multicolumn{1}{c|}{90.5}            \\
~~Invariant mass $> 10$ GeV                  & 46.5                 & \multicolumn{1}{c|}{47.6}          & 39.8                 & \multicolumn{1}{c|}{40.4}          & 85.9                & \multicolumn{1}{c|}{86.9}          & 89.9                &         \multicolumn{1}{c|}{90.5}  \\ 
\hline
\end{tabular}
}
\caption{Trackless-jet SR acceptance with full cutflow. With the EW-channel production and the electroweakinos decaying to light-flavor quarks.}
\label{tab:trackless_LF_acc_full}
\end{center}
\end{table}

\begin{table}[t]
\begin{center}
\begin{tabular}{c| c c c c}
\hline
$m(\tilde{\chi}^{0}_{1}), \tau(\tilde{\chi}^{0}_{1})$ & Full Sim. & Param.~Exp. & Param.~Ours & Non-closure \\
\hline
500 GeV, 0.1 ns & 31.1$\%$ & 28.1$\%$ & 34.6$\%$ & 11.3$\%$ \\
500 GeV, 1 ns & 14.3$\%$ & 14.3$\%$ & 24.9$\%$ & 74.1$\%$ \\
1300 GeV, 0.1 ns & 12.2$\%$ & 11.7$\%$ & 11.1$\%$ & -9.0$\%$ \\
1300 GeV, 1 ns & 8.3$\%$ & 7.9$\%$ & 11.7$\%$ & 41.0$\%$ \\
\hline
\end{tabular}
\caption{Trackless-jet SR $\epsilon$ including both the acceptances' and the parameterized efficiencies' effects. With the EW-channel production and the electroweakinos decaying to light-flavor quarks.}
\label{tab:trackless_LF_acc_eff}
\end{center}
\end{table}

Similarly, we show in Table~\ref{tab:trackless_LF_acc_full} and Table~\ref{tab:trackless_LF_acc_eff} the validation results for the Trackless-jet SR in the EW-production channel of the electroweakinos decaying to light-flavor quarks.
We come to similar conclusions that while for the acceptance cutflows we achieve very good validation, once the parameterized efficiencies are taken into account, only some benchmarks are well validated and the others show non-closure values up to about 70\%.

\begin{table}[t]
\begin{center}
\resizebox{\textwidth}{!}{%
\begin{tabular}{|l|cccccc}
\hline
                                             & \multicolumn{6}{c|}{Acceptance $[\%]$}                                                                                                                                 \\ \cline{2-7} 
\multicolumn{1}{|c|}{$m(\tilde{\chi}^0_1)$ [GeV]}                  & \multicolumn{2}{c|}{1500}                              & \multicolumn{2}{c|}{1500}                               & \multicolumn{2}{c|}{1500}                              \\
\multicolumn{1}{|c|}{$\tau(\tilde{\chi}^0_1)$ [ns]}                & \multicolumn{2}{c|}{0.032}                             & \multicolumn{2}{c|}{0.1}                              & \multicolumn{2}{c|}{1}                                \\ \hline
Selection                                    & Exp.                 & \multicolumn{1}{c|}{This work} & Exp.                 & \multicolumn{1}{c|}{This work} & Exp.                 & \multicolumn{1}{c|}{This work} \\
Jet selection                                & 84.7                 & \multicolumn{1}{c|}{82.5}          & 84.7                 & \multicolumn{1}{c|}{82.4}          & 84.7                 & \multicolumn{1}{c|}{82.4}          \\
Event has $\geq 1$ DV passing:               & \multicolumn{1}{l}{} & \multicolumn{1}{l|}{}          & \multicolumn{1}{l}{} & \multicolumn{1}{l|}{}          & \multicolumn{1}{l}{} & \multicolumn{1}{l|}{}          \\
~~$R_{xy}, |z|<300$ mm                       & 84.7                 & \multicolumn{1}{c|}{82.5}          & 84.7                & \multicolumn{1}{c|}{82.4}          & 80.1                 & \multicolumn{1}{c|}{78.0}          \\
~~$R_{xy}>4$ mm                              & 45.7                & \multicolumn{1}{c|}{46.9}          & 73.3                 & \multicolumn{1}{c|}{72.3}          & 78.4                 & \multicolumn{1}{c|}{76.5}          \\
~~$\geq 1$ track with $|d_0|>2$ mm           & 45.7                & \multicolumn{1}{c|}{46.9}          & 73.3                & \multicolumn{1}{c|}{72.3}          & 78.4                 & \multicolumn{1}{c|}{76.5}          \\
~~$n_{\text{selected decay products}}\geq 5$ & 45.7                 & \multicolumn{1}{c|}{46.9}          & 73.3                 & \multicolumn{1}{c|}{72.3}          & 78.4                 & \multicolumn{1}{c|}{76.5}          \\
~~Invariant mass $> 10$ GeV                  & 45.7                 & \multicolumn{1}{c|}{46.9}          & 73.3                 & \multicolumn{1}{c|}{72.3}          & 78.4                 & \multicolumn{1}{c|}{76.5}         \\
\hline
\end{tabular}
}
\caption{High-$p_T$-jet SR acceptance with full cutflow. With the EW-channel production and the electroweakinos decaying to heavy-flavor quarks. Note that for this scenario, the EW-production channel, instead of the strong channel, is considered, following the recasting material of Ref.~\cite{ATLAS:2023oti}.}
\label{tab:highpt_HF_acc_full}
\end{center}
\end{table}

\begin{table}[t]
\begin{center}
\begin{tabular}{c| c c c c}
\hline
$m(\tilde{\chi}^{0}_{1}), \tau(\tilde{\chi}^{0}_{1})$ & Full Sim. & Param.~Exp. & Param.~Ours & Non-closure \\
\hline
1500 GeV, 0.032 ns & 39.6$\%$ & 42.7$\%$ & 45.6$\%$ & 15.2$\%$ \\
1500 GeV, 0.1 ns & 57.7$\%$ & 62.7$\%$ & 68.9$\%$ & 19.4$\%$ \\
1500 GeV, 1 ns & 36.7$\%$ & 43.0$\%$ & 65.0$\%$ & 77.1$\%$ \\
\hline
\end{tabular}
\caption{High-$p_T$-jet SR $\epsilon$ including both the acceptances' and the parameterized efficiencies' effects. With the EW-channel production and the electroweakinos decaying to heavy-flavor quarks.}
\label{tab:highpt_HF_acc_eff}
\end{center}
\end{table}

\begin{table}[t]
\resizebox{\textwidth}{!}{%
\begin{tabular}{|l|cccccc}
\hline
                                             & \multicolumn{6}{c|}{Acceptance $[\%]$}                                                                                                                                 \\ \cline{2-7} 
\multicolumn{1}{|c|}{$m(\tilde{\chi}^0_1)$ [GeV]}                   & \multicolumn{2}{c|}{700}                              & \multicolumn{2}{c|}{700}                               & \multicolumn{2}{c|}{700}                              \\
\multicolumn{1}{|c|}{$\tau(\tilde{\chi}^0_1)$ [ns]}                & \multicolumn{2}{c|}{0.032}                             & \multicolumn{2}{c|}{0.1}                              & \multicolumn{2}{c|}{1}                                \\ \hline
Selection                                    & Exp.                 & \multicolumn{1}{c|}{This work} & Exp.                 & \multicolumn{1}{c|}{This work} & Exp.                 & \multicolumn{1}{c|}{This work} \\
Jet selection                                & 69.8                 & \multicolumn{1}{c|}{72.2}          & 74.1                 & \multicolumn{1}{c|}{72.2}          & 74.7                 & \multicolumn{1}{c|}{71.9}          \\
Event has $\geq 1$ DV passing:               & \multicolumn{1}{l}{} & \multicolumn{1}{l|}{}          & \multicolumn{1}{l}{} & \multicolumn{1}{l|}{}          & \multicolumn{1}{l}{} & \multicolumn{1}{l|}{}          \\
~~$R_{xy}, |z|<300$ mm                       & 69.8                 & \multicolumn{1}{c|}{72.2}          & 74.1               & \multicolumn{1}{c|}{72.2}          & 64.8                & \multicolumn{1}{c|}{62.4}          \\
~~$R_{xy}>4$ mm                              & 48.4                & \multicolumn{1}{c|}{48.5}          & 68.1                 & \multicolumn{1}{c|}{66.3}          & 62.9                 & \multicolumn{1}{c|}{60.9}          \\
~~$\geq 1$ track with $|d_0|>2$ mm           & 48.4                & \multicolumn{1}{c|}{48.5}          & 68.1                & \multicolumn{1}{c|}{66.3}          & 62.9                 & \multicolumn{1}{c|}{60.9}          \\
~~$n_{\text{selected decay products}}\geq 5$ & 48.4                 & \multicolumn{1}{c|}{48.5}          & 68.1                 & \multicolumn{1}{c|}{66.3}          & 62.9                 & \multicolumn{1}{c|}{60.9}          \\
~~Invariant mass $> 10$ GeV                  & 48.4                 & \multicolumn{1}{c|}{48.5}          & 68.1                 & \multicolumn{1}{c|}{66.3}          & 62.9                 & \multicolumn{1}{c|}{60.9}      \\
\hline
\end{tabular}
}
\caption{Trackless-jet SR acceptance with full cutflow. With the EW-channel production and the electroweakinos decaying to heavy-flavor quarks.}
\label{tab:trackless_HF_acc_full}
\end{table}

\begin{table}[t]
\begin{center}
\begin{tabular}{c| c c c c}
\hline
$m(\tilde{\chi}^{0}_{1}), \tau(\tilde{\chi}^{0}_{1})$ & Full Sim. & Param.~Exp. & Param.~Ours & Non-closure \\
\hline
700 GeV, 0.032 ns & 26.6$\%$ & 28.2$\%$ & 30.0$\%$ & 12.8$\%$ \\
700 GeV, 0.1 ns & 37.5$\%$ & 36.7$\%$ & 42.5$\%$ & 13.3$\%$ \\
700 GeV, 1 ns & 20.0$\%$ & 21.1$\%$ & 34.9$\%$ & 74.5$\%$ \\
\hline
\end{tabular}
\caption{Trackless-jet SR $\epsilon$ including both the acceptances' and the parameterized efficiencies' effects. With the EW-channel production and the electroweakinos decaying to heavy-flavor quarks.}
\label{tab:trackless_HF_acc_eff}
\end{center}
\end{table}

Moving to the benchmarks where the electroweakinos decay to heavy-flavor quarks, we present in Table~\ref{tab:highpt_HF_acc_full}, Table~\ref{tab:highpt_HF_acc_eff}, Table~\ref{tab:trackless_HF_acc_full}, and Table~\ref{tab:trackless_HF_acc_eff}, the corresponding validation results.
Here, only the EW-production channels are considered, and both the High-$p_T$-jet and Trackless-jet SRs are applied.
We reach a similar conclusion as in the previous benchmarks, that the acceptance-only cutflows agree very well with the experimental published results, but with the inclusion of the parameterized efficiencies we find relatively unsatisfactory comparisons with large non-closure values in some benchmarks.

Given the generally excellent validation results with the acceptance requirements only, we argue that we should have selected the correct set of events from the whole event samples.
Moreover, the event-level and vertex-level efficiencies are functions with input parameters that determine the acceptance-level cutflows.
Therefore, after excluding possible issues within coding itself, we do not find an explanation for the discrepancies observed once the parameterized efficiencies are included, and thus urge for further collaboration between the theorists and experimentalists in order to resolve the issue.
To aid this end, we have uploaded our code to the public LLP Recasting Repository~\cite{LLPrepo}, hoping it would be useful for other groups.
\section{Recast of single top+jets/ MET searches}\label{sec:recast_single_top}

In this appendix, we detail a simple recast for single-top events in terms of the ALP scenario studied here, following closely the procedures implemented in Ref.~\cite{Carmona:2022jid}.

The considered ALP scenario leads to large production rates of a single top quark associated with an ALP: $pp\to t a$.
Since the ALP in our study dominantly decays into jets, the process can mimic the SM processes and signatures of $pp\to t j$, $pp\to t+$ displaced jets or vertices, and $pp\to t+$ MET, depending on the lifetime of the ALP.
These searches can be complementary to the ATLAS DV+jets search which is the focus of this work, especially for promptly decaying or very long-lived ALPs.

The CMS search~\cite{CMS:2016uzc} focuses on a single leptonic top plus prompt jets with $tqg$ couplings with $q$ denoting the $u$ or $c$ quark and $g$ the SM gluon.
At least one jet should fail a $b$-tagging secondary-vertex algorithm~\cite{CMS:2017wtu} which selects jets in the range $0.01$ cm $< r < 2.5$ cm with $r$ labelling the transverse distance of the jet's production location to the collision point.
We assume that a slightly displaced ALP would behave in a similar way as the SM $B$-hadrons, and, hence, in order to recast the search, we require that jets be selected, or, equivalently, the ALPs decay, within one of the regions with 0 cm $<r<0.01$ cm and $2.5$ cm $<r < 2$ m, so that the ALP is probably accepted by the search.
Here, the distance 2 m is chosen~\cite{Carmona:2022jid} in accordance with the hadronic calorimeter's size.
This search can thus constrain the jet-like ALPs with relatively shorter lifetimes.
In addition, an ATLAS search reported in Ref.~\cite{ATLAS:2015iqc} targets a single-top channel with FCNC top quarks also via a gluon mediator.
The search requires exactly one jet and one lepton from the top quark, as well as MET.
Such a search can exclude certain parameter regions of our ALP scenario for longer lifetimes of the ALP which behaves as MET at the detector level.
For the recast of this search, we simply require that the ALP should decay outside $r=10$ m which roughly corresponds to the scale of the ATLAS detector's geometrical configurations.
Finally, upper bounds on new-physics contributions to the scattering cross sections at 13-TeV LHC of single top plus jets or MET processes have been obtained in Ref.~\cite{Goldouzian:2016mrt} considering the above searches: $\sigma_{tj}\sim 0.29$ pb and $\sigma_{t}\sim 0.10$ pb.
These bounds can be used to set limits on models of the ALPs coupled to the top quark.

We thus generate events of $pp\to t a$ at $\sqrt{s}=13$ TeV with MadGraph5 and apply an acceptance factor to the computed scattering cross sections according to the fiducial-volume regions specified above:
\begin{eqnarray}
    \epsilon^{\text{acc.}}_{\text{prompt}} &=& \sum_{i=1}^{N_{\text{MC}}} \Big( (1 - e^{\frac{-10^{-4}\text{ m}}{\beta_i^r\gamma_i c \tau_a}}    )   + ( e^{\frac{-2.5\times 10^{-2}\text{ m}}{\beta_i^r\gamma_i c \tau_a}}  - e^{\frac{-2\text{ m}}{\beta_i^r\gamma_i c \tau_a}}) \Big),\\
    \epsilon^{\text{acc.}}_{\text{MET}} &=& \sum_{i=1}^{N_{\text{MC}}} e^{\frac{-10\text{ m}}{\beta_i^r\gamma_i c \tau_a}},
\end{eqnarray}
where $N_\text{MC}=10000$ is the number of MC events we simulated, and $\beta_i^r$ and $\gamma_i$ are, respectively, the transverse speed and the boost factor of generated ALP in the $i$-th simulated event.
Requiring that the acceptance-weighted cross sections be below the upper limits $\sigma_{tj}$ and $\sigma_{t}$ allows to obtain bounds on the ALP couplings depending on its mass.

The results are shown in Fig.~\ref{fig:sensitivities_g_vs_g} for comparison with those obtained in this work.
The exclusion bounds from the top+jet (top+MET) search are shown in dotted (dot-dashed) line style.
We observe that the top+MET search is sensitive only to rather small values of the decay couplings corresponding to larger lifetimes of the ALP.
The sensitivity of the top+jet search to the production couplings is weakened at $\frac{g_{ii}}{\Lambda}\sim (2-6)\times 10^{-8}$ GeV$^{-1}$ depending on the ALP's mass, because the region $0.01$ cm $< r < 2.5$ cm is excluded from the fiducial volume, for masses except 90 GeV.
For $m_a=90$ GeV, both the production and decay of the ALP are dominated by the production couplings and as a result, the top+jet search's bounds are shown as a horizontal line.
See Sec.~\ref{sec:results} for more discussion.
In principle, the reinterpreted results of these single-top searches can also be plotted in Fig.~\ref{fig:sensitivities_g_vs_m} where the production couplings are fixed to be 150 or 1000 times the decay couplings.
However, it turns out that with these fixed relations, the top+MET search is insensitive to the scenario, and the top+jet search can only constrain the ALP couplings larger than those bounded by the ATLAS+jets search.
Therefore, we choose not to display these results in Fig.~\ref{fig:sensitivities_g_vs_m}.

\bibliographystyle{JHEP}
\bibliography{bib}

\end{document}